# Bayesian Searches and Quantum Oscillators

LLNL-CONF-764010

G. Chapline[1] and M. Otten[2]

3rd Workshop on Microwave Cavities and Detectors, April 2018

1. LLNL 2. ANL

A new LLNL Strategic Initiative is focused on developing improved methods for Bayesian inference when the input data depends on hidden parameters. Part of this effort involves investigating the idea of using an array of quantum oscillators (*viz* microwave cavities) as an analog computer for implementing Bayesian model selection. The practical motivations are twofold: 1) Bayesian model selection problems are often intractable using conventional digital computers, and 2) quantum information processing may allow detection of weak analog signals below the usual quantum noise threshold.

The working hypothesis for our current LLNL quantum Bayesian effort is that quantum dynamics of coupled harmonic oscillators may alleviate some of the difficulties encountered when trying to optimize Bayesian model selection problems; e.g. finding the optimum strategy for complex Bayesian search problems. This hypothesis already has a certain plausibility by virtue of the discovery at Google Brain [1] that neural network approaches to pattern recognition can in many situations of interest be represented in terms of manipulation of Gaussian processes (GPs); i.e. variables in a large dimensional vector space where all the variables of interest are represented using *iid* Gaussian probability densities. Since the quantum wave functions for harmonic oscillators involve Gaussian functions, it seems natural to surmise that the quantum dynamics of oscillator arrays might potentially be useful for finding weak signals in analog signals. Indeed this may be of some interest in connection with the problem of searching for axions with superconducting microwave cavities.

Impetus for this effort is provided by another LLNL LDRD project aimed at the development of a 4 qudit quantum computer as a user facility, where each qudit consists of the lowest quantized energy levels of a superconducting microwave cavity. If one interprets the lowest N levels of a microwave oscillator as representing N locations, and each location is paired with a qubit, then might imagine that such a system could be used to represent a "Monty Hall" type Bayesian search. In this simple type of a Bayesian search one is searching for an object concealed in one of N boxes, and the auxillary qubit can be used to represent the probability that the object is present or not at the paired location. For example, the case of the axion search one might identify each location with the axion frequency. This type of problem differs from the typical machine learning problem, where one is just trying to fit the explanations for a training set of observed data inputs with a parameterized interpolation model, in that there is an additional hidden "factor" that influences the input data – the location of the object. In this talk I describe some of the progress we have made towards emulating the probabilistic dynamics of a Bayesian search with an array of quantum oscillators.

In the 1990s it began to be appreciated the GPs representations for both input data and data interpretation models would provide a way of using Bayes' formula for posterior probabilities to construct analytic regression models, whose parameters are conditioned with past data, and which are able to provide an explanation for new data [2]. It has also been shown [3] that GP representations can be used facilitate the use of Bayesian methods for robotic control problems. The application of GP methods to control problems is based on the application of Bayes theorem at each stage of the search or control process to make probabilistic predictions for the future state of the system given a control decision. Actually there is a close relationship between Bayesian approaches to robotic control and Bayesian searches. However, whereas the only model parameters for GP-Bayesian control applications are the GP covariances and means, Bayesian searches involve a hidden parameter: the location of the object being sought. It seems natural to assume that the successes of GP-Bayesian techniques for solving robotic control problems can be extended to Bayesian searches, although this remains to be demonstrated.

Bayesian filter approach to search problems

A Bayesian search is defined by a sequence of control decisions $U_N \equiv \{x_{n+1} - x_n, n = 1,..,N-1\}$ leading to a sequence $X_N \equiv \{x_n, n=1,..,N\}$ of locations or system states that will be interrogated for the presence of a desired object or system state $x^*$. The interrogations result in a sequence of measurements $Y_M \equiv \{y_k, k =1,..M\}$ that are intended is to determine at each step k whether the desired object or state $x^*$ is present. In reality the result of each observation, $y_k$, only determines probabilistically whether $x_k$ is the desired state $x^*$. Based on the collection $Y_n$ of measurements gathered during the first n steps of a search and the prior posterioir probability density $p_{n-1}(x)$, one can estimate the posterior probability density at step n $p_n(x) \equiv p_n(x|Y_n, X_n, p_{n-1})$ that the desired object or state $x^*$ the "Bayes Filter" for predicting the future becomes :

$$p_n(x^* = x|Y_n, U_n, p_{n-1})$$
$$= \frac{P(Y_n|U_n, x^* = x, p_{n-1})p_{n-1}(x^* = x|Y_{n-1}, U_{n-1})}{P(Y_n|U_n)} \quad (1)$$

Since the results of using a Bayes Filter are expressed as probabilities, it is natural to inquire how much information has been gained about the location of the desired object after N decision steps. A natural measure of the information gathered after n-steps is the Shannon entropy

$$H(p_n) = -\int p_n(x) \log p_n(x) dx$$

where the log is base 2. The optimal search strategy can now be characterized as the problem of finding a policy for choosing the control sequence $\{U_n\}$ so that the Shannon entropy $H(p_N)$ is minimized after N-steps.

As shown in Ref [3] the use of Gaussian process representations for the variables $X_N$ and $U_N$ allows one to write down analytic expressions for the posterior probabilities $p_n(x^* = x)$ given a set of training data, which in turn allows one to make incremental improvements in the cost function $H(p_N)$. Unfortunately when many input data streams or hidden variables are involved evaluating the Bayes filter predictions can easily become intractable - even with the use of Markov chain Monte Carlo sampling techniques. As it happens there is a neural network-like architecture, the Helmholtz machine, and associated algorithm, the wake-sleep algorithm, for choosing the hidden parameters that minimize the Shannon entropy [4]. Unfortunately, in contrast with the gradient descent algorithm that can be used to find the parameters (means and variances) of a GP [3], use of the wake-sleep algorithm to find optimal hidden parameters with existing formulations of the Helmholtz machine, which use the binary spin nodes usually used for neural networks, is typically an intractable problem. Whether there is a GP version of the classical Helmholtz machine would be useful is an area of active research. On the other hand, we already have evidence that the quantum dynamics of interacting qubits might provide a framework for realizing the Helmholtz machine approach to finding hidden variable models for input data.

Self-organization approach to model selection
Our original inspiration for the idea that the quantum dynamics of oscillators might be useful for solving Bayesian inference problems with hidden variables was provided the observation that the Durbin-Willshaw elastic net method [5] (that actually predated the Helmholtz machine) for solving the traveling salesman problem (TSP) (cf. Fig 1) might be reformulated as quantum dynamics problem [6]. As the name suggests the elastic net approach to solving the traveling salesman problem involves replacing the path of the "salesman" by a strings of points – the square points in Fig. 1 - connected by springs, while the connection of these points to the cities that are to be visited – the round points in Fig. 1 - reflects the likelihood that a chosen point represents the visit to a particular city as part of the minimum path. It is almost obvious way that this setup can be translated to method for solving Bayesian search problems by representing the input data $\{Y_i\}$ as the positions of the round points in Fig.1, while the positions of the square points are parameters $w(x_n)$ defining a model for the path traveled. The string running through the square nodes plays the role of MacKay's regression model $Z(x)$ for input data. In the limit of weak coupling one can describe the Bayesian regression of the interpolation model as the quantum motion of the string variables string acted on by a stochastic classical field arising from the stochastic nature of the distances $d(i,\mu)$ he stochastic nature of the influence of the oscillators attached to the cities provides the dissipation needed to minimize the information cost for representing the positions of the nodes. The net result is the prior distribution for the feature variables has the form

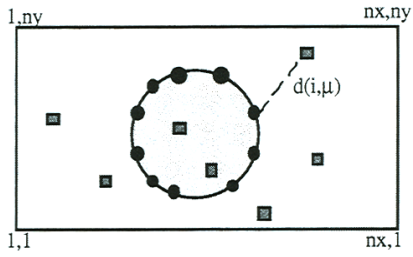

1. Elastic net solution for the traveling salesman problem

$$P(\{w(r_n)\}) = \prod_n \exp\left(-\frac{\alpha}{2}(\frac{\partial w}{\partial r_\alpha})^2\right) \qquad (2)$$

This "string action" plays essentially the same role as the prior distribution for the parameters for the generative model for input data used in the "sleep phase" of the Helmholtz machine [Hinton et al 1995]. In the Durbin-Willshaw method the data model represented the square points is adapted to the data by introducing in addition to the prior model eq. 5 a cost function for the difference between the model and observations of the form:

$$E_C[\{w(r_n)\}] = -\frac{\beta}{2}\sum_n \log\left(\sum_m [\exp - (|w(r_n) - \phi_m|^2)]\right) \qquad (3)$$

This cost function causes the feature vectors to be attracted to the to the "true" values for the model parameters, and corresponds to a likelihood predictive model of the form.

$$P(\phi_i|\{w(r_n)\}) = \prod_{n=1}^{N} exp\left(-\frac{\beta}{2}|w(r_n) - \phi_i|^2\right) \quad (4)$$

It is perhaps worth noting that from the point of view of quantum mechainics the Durbin-Willshaw attraction between the string nodes repesenting the locations of the cities and the fixed nodes representing the ciries corresponds to attracting the Gaussian wave funtions for the model oscillators to the ground state wave functions for particles bound to the city locations by a harmonic potential. We can now see how a quantum version of the Durbin-Willshaw network might be able to deal with optimizing the type of hidden parameter that occurs in the Monty Hall problem. The hidden variables in the traveling salesman problem can be thought of as variables $\lambda_i^n$ that link a particular point of the string with a particular city. It is fairly obvious from the structure of the Durbin-Willshaw energy function that the optimal values for these variable emerge automatically as the string relaxes to the shortest path threading the cities. In the quantum vesion of the Durbin-Willshaw network the string length becomes the action function for the quantum oscillators representing the string. Evidently then the Durban-Willshaw self-organizing net method for solving the traveling salesman problem (TSP) is essentially equivalent to taking the classical limit of a Feynman path integral for a system of points linked by quantum oscillators that are acted on by a stochastic external field.

Quantum dynamics for an oscillator coupled to qubits

The density matrix propagator for a quantum harmonic oscillator subject to an environment consisting of a classical noisy signal $f(t)$ random classical force is :

$$J = \iint e^{i\{S[x(t)]-S[y(t)]\}/\hbar} \Phi[x(t) - x'(t)] Dx(t) Dy(t) \quad (5)$$

where $S[x(t)]$ is the classical action for an isolated oscillator with mass m and angular frequency $\omega_0$,

$$\iint e^{i\{S[x(t)]-S[x'(t)]\}/\hbar} = \iint exp\left(\frac{im}{2\hbar}(\dot{x}^2 - \omega_0^2 x^2 - \dot{y}^2 + \omega_0^2 y^2)\right),$$

and $\Phi[k(t)]$ is the functional Fourier transform of the probability density functional for the noisy signal $f(t)$:

$$\Phi[k(t)] = \int e^{i\int k(t)f(t)dt} P[f(t)] Df(t) \quad (6)$$

For a Gaussian process signal one has

$$P[f(t)] = exp\left\{-\frac{1}{2}\iint (f(t) - \bar{f}(t))A^{-1}(t,t')(f(t') - \bar{f}(t'))dt dt'\right\}$$

where A is the autocorrelation function for the signal. If instead of a classical noise signal $f(t)$ the free oscillator is coupled to a quantum environment then the real classical functional $\Phi$ is replaced by an complex valued influence functional F{[x(t), x'(t')]}. The resulting theory of quantum noise, due independently to Feynman and Keldysh, implies that the density matrix obeys a non-Markovian master equation, which in the case of a superconducting cavity coupled to an array of qubits has the form:

$$\frac{\partial \rho}{\partial t} = \left(\frac{i\hbar}{2L}\left(\frac{\partial^2 \rho}{\partial Q^2} + \frac{\partial^2 \rho}{\partial Q'^2}\right) + \left(\frac{iL\omega_0^2}{2\hbar}(Q^2 - Q'^2)\right)\right.$$
$$\left. -\frac{C^2 \ln \Delta_{max}/\Delta_{min}}{\hbar^2 \Delta_{max}}[Q(t) - Q'(t)]\sum_{t_i=-\infty}^{t_i=t}\int_{t_i-\tau}^{t_i}(Q(s) - Q'(s))ds\right)\rho$$

(7)

where $\Delta_{min}$ and $\Delta_{max}$ are the minimum and maximum qubit level splittings. As first showed by Caldiera and Legget the nature of the density matrix evolution implied by (7) is that the forward and backward trajectories, $Q(t)$ and $Q'(t)$, converge with time. It is of course encouraging that this "wake-sleep" like behavior is just what we are looking for if we wish the behavior of the quantum oscillators representing models for either input data or data models to converge as indicated in Fig. 1. There is a Hamiltonian formulation of (7) which is more easily adapted to the qudit case,

$$H = \hbar\omega_r(a^\dagger a + \frac{1}{2}) + \sum_j[\frac{1}{2}\hbar\Delta_j\sigma_j^z + ig\sigma_j^y(a^\dagger - a)]$$

$$\dot{a} = -i\omega_0 - i\sum_j \frac{g_i}{\hbar}(\sigma_j^+ e^{-i\Delta(t-t_0)} - \sigma_j^- e^{i\Delta(t-t_0)}) + F(t)$$

(8)

$$F(t) = -\sum_j \frac{g_j^2}{\hbar^2}\int_{t_0}^{t}\sin(t-t')\sigma_j^z(t')[a^+(t') - a(t')]dt'$$

Because the "starting" times" $t_0$ for the episodes of coherent evolution for each qubit are randomly distributed the influence function F can now be regarded as a random function of time whose fluctuations can be measured by the autocorrelation function for the oscillator amplitude.

$$F(t) \approx \exp-\sum_j \frac{g_j^2}{\hbar^2}\left(\int_{t_{j0}}^{t}\int_{t_{j0}}^{t'}\left[e^{-i\Delta_j(t-t')}x(t')-e^{i\Delta_j(t-t')}y(t')\right]\left[x(t)-y(t)\right]dt'\,dt\right)$$

Thus it appears that the quantum dynamics of oscillators coupled to qubits does in fact provide a kind of quantum analog of the GP representations that have been found to be useful in the classical Bayesian analyses of regression and robotic control problems. Of course, in practice, one would have to work with qudits rather than the full Hilbert space of an oscillator as is implied in Eq. (7), and this is what we are currently pursuing.

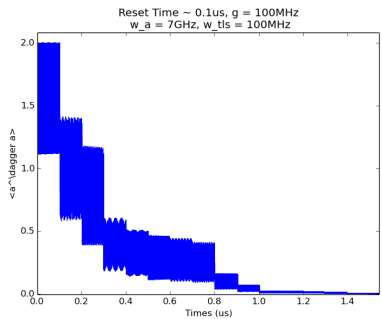

Fig 2. Decay of a qudit due to coupling to a qudit that is sporadically reset to |0>.

One important issue is how to represent the measurements that are in the inputs for a Bayesian search. We have already carried out simulations where at each instant of time, the density matrix is either left unchanged or reset to zero at a specified rate. The result that the initial excitation of the qudit decays with time as shown in Fig 2. One interesting aspect of this calculation is that the qudit decay occurs even if the coupling of the qubit to the qudit is so weak that it isn't excited to its excited state.

This work was performed under the auspices of the U.S. Department of Energy by Lawrence Livermore National Laboratory under Contract DE-AC52-07NA27344 and was supported by the LLNL-LDRD Program under Project No. 16-ERD-013.